\documentclass[prd,aps,amssymb,nofootinbib,showpacs,preprint]{revtex4}
\usepackage{amsmath,amsfonts}
\usepackage{textcomp,color}
\usepackage{hyperref}
\usepackage{graphics,graphicx}
\usepackage{epstopdf}
\usepackage{ulem}
\usepackage{overpic}

\begin{document}
	\title{Cosmology and matter induced branes }
	\author{\firstname{Sergey G.}~\surname{Rubin}}\email{sergeirubin@list.ru} \affiliation{National Research Nuclear University MEPhI (Moscow Engineering Physics Institute),\\ 115409, Kashirskoe shosse 31, Moscow, Russia}
	\affiliation{N.~I.~Lobachevsky Institute of Mathematics and Mechanics, Kazan  Federal  University, \\ 420008,   
		Kremlevskaya  street  18,  Kazan,  Russia} 
	\date{June 2019}
	\begin{abstract}
The extra space paradigm plays a significant role in modern physics and cosmology as a specific case. In this review, the relation between the main cosmological parameters -  the Planck mass and the Cosmological constants - and a metric of extra space is discussed. Matter distribution inside extra space and its effect on the 4-dimensional observational parameters is of particular interest.
The ways to solve the Fine-tuning problem and the Hierarchy problem are analyzed.    
	\end{abstract}
	
	\maketitle
	
	\section{Introduction}
	The Cosmology encompasses physical laws at all distances. The impressive interpenetration of microphysics and physics at extremely large distances has been noticed some time ago \cite{1981RvMP...53....1D} and is discussed up to now \cite{KhlopovRubin}.  The origin of the physical parameters like masses and coupling constants is the matter of future theory, but there are two general facts that are worth discussing. The first one relates to the smallness of all observable parameters as compared to the Planck mass (Hierarchy problem). The second one is known as the Fine-tuning problem \cite{2007unmu.book..231D,Page:2003zm}. The cosmological constant smallness is its most amazing illustration. Even more serious problem concerns the Anthropic observational fact - the increasing of the cosmological constant value in several times leads to a serious variation of our Universe structure so that we, observers would not exist. This is the particular case the general fact of the Fine-tuning of our Universe.

	The Planck mass and the cosmological constant seem to be "more fundamental" and important for different cosmological research. These two parameters and their dependence on an extra space structure are used throughout the text as the illustration of the discussion.

	\textbf{The Planck mass}
	is one of the natural units introduced by Max Planck. It is connected to the Newton constant $G_N$ as $M_P=\sqrt{8\pi/G_N}, \, (h=c=1)$. Its value is known with poor accuracy that is the reason for a variety of speculation on its origin and time variation. For example, the authors of the paper \cite{vandeBruck:2015gsa} consider the Planck mass depending on a scalar field that tends to constant shortly after the inflation. The Planck mass is in many orders of magnitude greater the electroweak scale. This puzzle is known as the Hierarchy problem is not clarified yet.

	The standard Einstein-Hilbert gravity with the Lambda term $\Lambda$ is described by the action
	\begin{equation}\label{EH}
	S=\int d^4 x \left[\frac{1}{16\pi G_N}R-\Lambda\right], M_P^2 = \frac{1}{8\pi G_N}.
	\end{equation}
	Here $R$ is the Ricci scalar of our 4-dim space and $\Lambda$ is the cosmological constant. The Newton constant does not vary with time by definition. There are research studied its possible slow time variation \cite{Bronnikov:2003rf}, but we do not discuss this direction here.

	Action \eqref{EH} is firmly confirmed at the energies lower than $\sim 10$TeV. A variety of modified gravitational actions are studied at higher scales,  \cite{Gogberashvili:1998iu,Lyakhova:2018zsr}.  The necessary condition for all of them is the reduction to the standard form \eqref{EH} at the low energies. We will see how does it work for multidimensional models.
	
	The vacuum energy is connected to  \textbf{the cosmological constant} (the $\Lambda$ term) which is included now into the Standard Cosmological Model. The substantial review may be found in \cite{Sahni:1999gb}. Common opinion nowadays is that this energy is positive and extremely small which leads to several consequences.  First, our Universe is expanding with acceleration; second, the modern horizon is shrinking with time and third, the large scale structure was formed under the strong influence of the positive energy density uniformly distributed in the space. 
	
	There are two riddles related to the $\Lambda$ term. One of them is its smallness which can not be explained by now. Another one is the coincidence problem: the energy density of usual matter distributed at large scales is very close to the vacuum energy density nowadays. 
	
	A substantial amount of attempts to solve the problems mentioned above is based on the idea of the extra dimensions \cite{1999NuPhB.537...47D}.

	\section{The extra dimensions}
	
	In modern physics, the idea of extra dimensions is used for explanation of a variety of phenomena. It is applied for the elaboration of physics beyond the Standard Model, cosmological scenarios including inflationary models and the origin of the dark component of the Universe (dark matter and energy), the number of fermion generations and so on. Gradually, this direction becomes the main element for a future theory. Sometimes extra dimensions are endowed by scalar fields and form fields to stabilize their metric. There are models where the Cazimir effect is attracted for the same reason \cite{Fischbach:2009ud}, \cite{2018GrCo...24..154B}. Our experience indicates that we live in the 4-dimensional world so that a mechanism to hide the extra dimensions is a necessary element of each model.

	Let us describe some models focusing on the Planck mass, the Lambda term and the two related problems mentioned at the beginning of the Introduction.
	
	\subsection{From D dimensions to 4 dimensions. General remark}\label{remark}
	
	Necessary element of all multidimensional models is a reduction of a D-dimensional action to an effective 4-dimensional form:
	\begin{equation}\label{reduction}
	\int d^DX\sqrt{|g_D|} L_D (\alpha_D,g_D)\rightarrow \int d^4x\sqrt{|g_4|} L_4 (\alpha_4,g_4)
	\end{equation}Here $\alpha_D$ is a set of $D-$dim parameters and $\alpha_4$ is general notation for the observable parameters like masses, coupling constant. The Planck mass $M_P$ and the cosmological constant $\Lambda$ are of particular interest. Fields dependence is assumed but not shown explicitly in \eqref{reduction}. An extra dimensional part $g_{extra}$ of the $D-$dimensional metric $g_D$ is hidden in the 4-dimensional parameters $\alpha_4$ so that 
	\begin{equation}\label{extra}
	\alpha_4 =\alpha_4(\alpha_D, g_{extra}).
	\end{equation} 
	
	A variety of observational parameters can be obtained by a variation of the extra space metrics. This remark is important for further discussion. Evidently, formula \eqref{extra} relates to the Planck mass and the cosmological constant as well.

	\subsection{The Planck mass and the extra space structure}
	
	\subsection*{Kaluza-Klein model}
	The action for this model has the form
	\begin{equation}\label{KK}
	S_g=\int d^4 xd^ny\sqrt{|g|} \left[\frac{m_D^{D-2}}{2}R-\Lambda\right].
	\end{equation}
	After integration out the extra dimensional coordinates $y$ we obtain effective action \eqref{EH} with the Planck mass related to the $D$-dimensional Planck mass $m_D$. The relation
	\begin{equation}\label{MPKK}
	M_P^2 = m_D^{D-2}v_n
	\end{equation}
	is the simplest realization of formula \eqref{extra}. Here $v_n$ stands for an extra space volume, $n=D-4$.
	Classical behavior of the system is possible if the inequality $v_n^{1/n}>1/m_D$ and hence $m_D<M_P$ take place. The latter is usually assumed, but it is optional, as we will see later.
	It is supposed that the fluctuations of known fields within the extra dimensions are very massive and can not be excited in the course of low energy processes.

	\subsection*{Hyperbolic extra dimensions}
	The conclusion on the size of the extra dimensions made above assumed the constant positive curvature of this extra space. The more encouraging result can be obtained if we attract a constant negative metric. In this case,  there is no rigid connection between the Ricci scalar and a characteristic size $L$ of a compact hyperbolic space which is the significant feature of such spaces \cite{2002PhRvD..66d5029N}. The volume of such manifold is
	
	\begin{equation}
	v_n=r_c^n e^{\alpha}, \quad \alpha \simeq (n-1)L/r_c
	\end{equation}
	where $r_c$ is the curvature radius and $L$ is the size of extra space which is not a Lagrangian parameter but an accidental value. The Planck mass exponentially depends on the independent linear size
	\begin{equation}\label{MP2}
	M^2 _{P}=m_D ^{n+2}v_n\simeq m_D ^{n+2}r_c^ne^{(n-1)L/r_c}
	\end{equation}
	and hence can be sufficiently large even if the Lagrangian parameters are fixed.
	
	\subsection*{f(R) theories}\label{fR}
	
	Nowadays, the $f(R)$ theories of gravity or more generally the theories with higher derivatives are widely used as the tool for the theoretical research. The interest in $f(R)$ theories is motivated by inflationary scenarios starting with the
	pioneering work of Starobinsky \cite{Starobinsky:1980te}.
	A number of viable $f(R)$ models in 4-dim space that satisfies the observable constraints are discussed in Refs. \cite{2014JCAP...01..008B,2007PhLB..651..224N,Sokolowski:2007rd}.
	
	$\Lambda CDM$ model successfully explains main part of the observational data. Nevertheless, it fails to describe such important phenomena as the dark matter and dark energy. Modern colliders have not detected the dark matter particles and there is no way to detect the dark energy density that is uniformly distributed in the space. The $f(R)$ models are suitable for explaining these two problems, as well as the phenomena of baryogenesis and inflation. Significant discussion on this subject can be found in \cite{Nojiri_2011, Nojiri_2017}.

	Consider the gravity with higher order derivatives and the action in the form,
	\begin{eqnarray}\label{act1}
	&& S=\frac{m_D ^{D-2}}{2}\int d^{D}Z \sqrt{|g_D|}f(R)
	\end{eqnarray} 
	The metric is assumed to be the direct product $M_4\times V_n$ of the 4-dim space $M_4$ and $n$-dim compact space $V_n$
	\begin{equation}\label{metric}
	ds^2 =g_{6,AB}dz^A dz^B = \eta_{4,\mu\nu}dx^{\mu}dx^{\mu} + g_{n,ab}(y)dy^a dy^b.
	\end{equation} 
	Here $\eta_{4,\mu\nu}$ is the Minkowski metric of the manifolds $M_4$ and $g_{n,ab}(y)$ is metric of the manifolds $V_n$.  $x$ and $y$ are the coordinates of the subspaces $M_4$ and $V_n$. We will refer to 4-dim space $M_4$ and $n$-dim compact space $V_n$ as the main space and an extra space respectively. The metric has the signature (+ - - - ...), the Greek indexes $\mu, \nu =0,1,2,3$ refer to 4-dimensional coordinates. Latin indexes run over $a,b = 4, 5,...$.
	
	According to \eqref{metric}, the Ricci scalar represents a simple sum of the Ricci scalar of the main space and the Ricci scalar of extra space
	\begin{equation}
	R=R_4 + R_n . 
	\end{equation}
	In this subsection, the extra space is assumed to be maximally symmetric so that its Ricci scalar $R_n=const$.
	In the following, natural inequality
	\begin{equation}\label{ll}
	R_4 \ll R_n
	\end{equation}
	is assumed. This suggestion looks natural for the extra space size $L_n < 10^{-18}$ cm if one compares it to the Schwarzschild radius $L_n \ll r_g \sim 10^6$cm of stellar mass black hole
	where the largest curvature in the modern Universe exists. Below we follow the method developed in \cite{Bronnikov:2005iz} 
	
	Using inequality \eqref{ll} the Taylor expansion of $f(R)$ in  Eq. \ref{act1} gives
	\begin{eqnarray}\label{act2}
	&& S= \frac{m_D ^{D-2}}{2}\int d^4 x d^n y \sqrt{|g_4(x)|} \sqrt{|g_n(y)|} f(R_4 + R_n )\\
	&& \simeq \frac{m_D ^{D-2}}{2}\int d^4 x d^n y\sqrt{|g_4(x)|} \sqrt{|g_n(y)|}[ R_4(x) f' (R_n) + f(R_n)]  \nonumber 
	\end{eqnarray}
	
	The prime denotes the derivation of function on its argument. Thus, $f'(R)$ stands for $df/dR$ in the formula written above.
	In this paper a stationary and uniform distribution of the matter fields in the 4-dimensional part of our Universe is assumed. Comparison of the second line in expression \eqref{act2} with the Einstein-Hilbert action
	\begin{equation}
	S_{EH}=\frac{M^2_{P}}{2} \int d^4x  \sqrt{|g(x)|}(R-2\Lambda)
	\end{equation}
	gives the expression  
	\begin{equation}\label{MPl}
	M^2 _{P}=m_{D} ^{D-2}v_n f' (R_n)
	\end{equation}
	for the Planck mass. Here $v_n$ is the volume of the extra space. The term
	\begin{equation}\label{Lambda0}
	\Lambda \equiv -\frac{m_D ^{D-2}}{2M^2 _{Pl}}v_n f(R_n)
	\end{equation}
	represents the cosmological $\Lambda$ term. Both the Planck mass and the $\Lambda$ term depends on a function $f(R)$. According to \eqref{MPl}, the Planck mass could be smaller than $D$-dim Planck mass, $M_P<m_D$ for specific functions $f$. This leads to nontrivial consequences. For example, the classical 4-dim observer is limited by the smallest scale $l_{quantum}\sim 1/M_P$. At high energy scales, an observer "feel"\ extra dimensions and hence the classical behaviour starts at the scale $l_{D,quantum}\sim 1/m_D$ which can be much smaller then $10^{-43}$cm, the standard Planck scale. This question deserves a separate discussion in future.

	\subsection*{Brane models}
	
	The first brane models have appeared two decades ago \cite{ArkaniHamed:1998rs,Gogberashvili:1998iu,1999PhRvL..83.3370R,2012arXiv1201.0614C}, see also nice review \cite{Shifman:2009df}, though the very first idea was declared in 1983 \cite{Rubakov:1983bb} where it was proposed that we are living in the  4-dimensional manifold that is immersed in a manifold of larger dimensions.

	A large but compact extra dimension was invented by Nima Arkani-Hamed, Savas Dimopoulos, and Gia Dvali \cite{ArkaniHamed:1998rs} (the ADD model). In this approach, the fields of the Standard Model are confined to a four-dimensional membrane, while gravity propagates in several additional spatial dimensions. The Planck mass relates to the extra space radius as
	\begin{equation}
	M_P=(2\pi R)^{n/2}m_D^{\frac{n+2}{2}}
	\end{equation}
	
	The Randall-Sundrum model \cite{1999PhRvL..83.3370R} is based on the 1-dim extra space representing $S^1/Z_2$ orbifold. Here $S^1$ is the circle and $Z_2$ is the multiplicative group $\{ -1, 1\}$. Two 3-dim branes are attached to two fixed points with coordinates $y=0$ and $y=L$.
	The 5-dim action is described by the expression
	\begin{equation}
	S_{RS}=S_g - \int d^4x\int dy\sqrt{|g_5|}\sigma_1 \delta(y) - \int d^4x\int dy\sqrt{|g_5|}\sigma_2 \delta(L-y).
	\end{equation}
	The first term is represented in \eqref{KK}. The second and the third terms describe the branes with the constant tensions $\sigma_1$ and $\sigma_2$.  
	
	The metric of the model describes the warped space with interval
	\begin{equation}
	ds^2 = e^{-2A(y)}\eta_{\mu\nu}dx^{\mu}dx^{\nu}-dy^2 
	\end{equation}
	This model solves the Hierarchy problem, the price of which is the connection of the Lagrangian parameters
	\begin{equation}\label{RSconnec}
	\sigma_1 = -\sigma_2 = \sqrt{-6m_D^3\Lambda}. (\Lambda<0)
	\end{equation}
	The 4-dim Planck mass is expressed in terms the model parameters as
	\begin{equation}\label{MPRS}
	M_P^2=\frac{1}{2k} (1-e^{-2kL})m_D^3, \quad k=\sqrt{\frac{-\Lambda}{6m_D^3}}.
	\end{equation}
	Evidentely, the 4-dim Planck mass $M_P$ and its 5-dim analog $m_D$ are of the same order of the magnitude for not very small value of the parameter $k$ and $L$.
	
	One can conclude that a solution to the Hierarchy problem looks solvable and the extra-dimensional paradigm is the important idea allowing the progress in this direction. In general, the brane idea represents a powerful tool to solve deep questions of modern physics. For example, the large value of the Planck mass as compared to the electro-weak scale can be justified. 
	
	The Fine-tuning problem remains unsolved yet. The fact of the fine-tuning is supported by a lot of examples \cite{2007unmu.book..231D,Bauer:2010wj}. There are many attempts to solve each problem separately. In the paper  \cite{Krause:2000uj} warped geometry is used to the solution of the small cosmological constant problem. The hybrid inflation \cite{2002PhRvD..65j5022G} was developed to avoid the smallness of the inflaton mass. The paper \cite{1999PhRvL..83.3370R} describes the way to solve the smallness of the Gravitational constant. Nevertheless, all of them suffer the fine-tuning of Lagrangian parameters. We devote the following discussion to this subject.
	
	\subsection{Brane as a clump of matter?}
	The first brane models postulated the extra space metric and 3-dim spaces (branes) that are attached to their critical points. The modern trend consists of involving thick branes into consideration which are soliton-like solutions extended in extra coordinates. To build such solutions, the scalar field potential with several vacua states \cite{Peyravi:2015bra} is usually proposed. The one-dimensional kinks are studied for a long time and represent a substantial ground for the branes construction.
	
	The serious shortage of the approach mentioned above consists of a firm connection of model parameters and the effective low energy parameters \eqref{extra}. Even if a model including extra dimensions is able to solve the Hierarchy Problem, the Fine-tuning enigma is still far from resolution.  The problem is simply translated from the observable parameters to parameters of a specific model.
	
	An important feature of branes is their ability to concentrate the matter nearby. But what is the effect of matter on the very structure of brane? This subject is studied below. The encouraging analogy is that a gravitating substance can experience the Jeans instability, as we know from four-dimensional physics. One may expect the same effect in the extra space which should lead to the brane formation.
	
	Here we discuss the new mechanism of the branes construction which was revealed in \cite{Rubin:2015pqa}. A complicated form of the scalar field potential is not necessary for it is known that the scalar field with the potential $V(\phi)\propto \phi^2$ experiences the gravitational instability  \cite{Khlopov:1985jw}. In analogy with the 4-dim case, the scalar field could form stable clumps within the extra dimensions due to the gravitational interaction. This subject has been also studied in \cite{Gani:2014lka}, \cite{Rubin:2014ffa}, \cite{2017JCAP...10..001B} and we shortly discuss it in next Section.
	
	The solution describing the brane depends on an initial amount of matter and hence such solutions form a continuous set. This property is extremely important for the discussion of the Fine-tuning problem and the Lambda term problem as a particular case.
	
	\section{Matter induced branes}

This section is the most important part of the research and it seems necessary to outline the idea. There is a well-known fact, that fields and the space-time metric experience quantum fluctuations in the very early Universe. 
According to the ideology of chaotic inflation, the space consists of a variety of causally disconnected domains filled by the scalar field.  The energy density of the scalar field within a volume under the horizon is an accidental value that varies continuously in wide but an uncertain range. The Multiverse is a set of such space domains (universes) within the horizons that were born at the sub-Planckian scale. The visible Universe represents a small subset of such domains. The fluctuations under the visible horizon are responsible for the observable large scale structure of the Universe.
	
Higher-dimensional inflation is also the subject of interest, see e.g. \cite{1991PhRvD..43..995H}. Relying on the 4-dimensional case, we suppose that the quantum fluctuations lead to the same consequences - the energy density varies accidentally in the D-dimensional world. That is, a scalar field $\phi(x,y)$ is different in different space domains (universes) which form the Multiverse. We know that the energy density evolves into local objects like galaxies and stars under the influence of the gravity in each 4-dimensional domain. In this paper, we suppose that the similar processes proceed also in the extra dimensions described by the $y$ coordinates. More definitely, it is supposed that the scalar field can be localized within the extra dimensions under the influence of the gravity. If that is true, the extra space metric is also an accidental value specific for each space domain. At the same time, the physical parameters at low energies depend on the extra-dimensional metric, see discussion in \ref{remark}. We come to the conclusion that the Multiverse consists of different universes with accidental physical parameters.
	
In this article, we examine whether static non-trivial metrics of the extra dimensions do exist. To this end, we suppose that all fields and metric are static, uniformly distributed in our 4-dimensional space and are heterogeneous in the extra space. This means that we limit ourselves by the scalar fields $\phi(y)$ and its energy density $\rho_{scalar}(y)\sim T_{00}(y)$ depending only on the extra-dimensional coordinates $y$. The contribution of the scalar field energy density  $\rho_{scalar}(y)$ to the $\Lambda$-term must be compensated by term containing the function $f(R)$, to obtain the observed value of the cosmological constant, see expression \eqref{Lambda2}. This point is checked at the final step.
	
	There are two questions to be clarified. The first one is how to find energy density within the extra dimensions. The second one is how to choose additional conditions that are necessary to solve the differential equations  \eqref{eqn1}, \eqref{eqn2} for unknown functions - the scalar field $\rho_{scalar}(y)$ and metric $g_{AB}(y)$. Evidently, these two questions are tightly connected. If we choose appropriate additional conditions we can find a solution to equations \eqref{eqn1}, \eqref{eqn2}, the knowledge of which allows us to calculate the energy density. This means that there is one to one correspondence between them: choice of additional conditions fixes the energy density and vice versa. Therefore, different energy density $\rho_{scalar}(y)$ in the space domains (universes) means that the additional conditions are also different there.

According to our numerical calculations, the deformation of the extra space metrics is concentrated near a critical point $\theta=0$ forming a brane as can be seen in Fig.~\ref{MetricVsMatter}. One can conclude that the form of the brane varies depending on the additional conditions \eqref{ad1}, \eqref{ad2}. This means that the branes also are accidental functions for each volume under the horizon and hence varies continuously depending on its position in the Multiverse.

Let us study the subject in more detail.

	\subsection{Matter distribution within extra space}
	
	This section discusses extra dimensions filled with an ordinary scalar field, which is accepted as the representative of matter. It is assumed that its potential has a single minimum.
	Solutions to the system of equations will indicate that the distribution of the scalar field has a critical point. As will be seen, the back reaction of the scalar field significantly affects the extra metric, forming non-trivial static configurations.
	
	Let us come back to the action \eqref{act1} with a scalar field $\phi$ 
	\begin{eqnarray}\label{actfL}
	S=\frac{m_D^{D-2}}{2}\int d^{D}z \sqrt{|g_D|}\left[f(R)+L_m\right]  
	\end{eqnarray} 
	where the function
	\begin{equation}
f(R)=aR^2 + bR +c
	\end{equation}
	is chosen in the simplest but nontrivial form
	and
	\begin{eqnarray}\label{act00}
	&&L_m=\left[\frac12 \partial_{A} \phi g_{D}^{AB}\partial_{B} \phi - \frac{m_{\phi}^2}{2}\phi^2 \right].
	\end{eqnarray}
	The corresponding equations of motion are as follows
	\begin{equation}\label{eqn}
	R_{AB} f' -\frac{1}{2}f(R)g_{AB} 
	- \nabla_A\nabla_B f_R + g_{AB} \square f' = \frac{1}{m_D^{D-2}}T_{AB}.
	\end{equation}
	Here $A=(\mu,a), \mu=1,2,3,4,\, a=5,6,..,D$, $\square$ stands by the d'Alembert operator
	\begin{equation}
	\square =\square_D=\frac{1}{\sqrt{|g_D|}}\partial_A ( {g_D}^{AB}\sqrt{|g_D|}\partial_B),\quad A,B=1,2,..,D
	\end{equation}
	Evidently, there is a continuum set of solutions to system \eqref{eqn} of the differential equations depending on additional conditions. Maximally symmetric extra spaces represent a small subset of this continuum set. One of the reasons to choose this particular case has been discussed in  \cite{Kirillov:2012gy}. As was shown there, the entropy outflow from the extra space into the large 3-dimensional space of our Universe leads to the maximally symmetric extra space at the final state.
	
	Essentially new element changing the situation is the matter (the scalar field) inclusion into the consideration. System \eqref{eqn} contains equation for the scalar field
	\begin{equation}\label{scalareq}
	\square_D \phi =- V'(\phi)
	\end{equation}
	Let us consider the class of the homogeneous in 4-dim space solutions $\phi(x,y)=\Phi(y)$ and suppose that the potential possesses unique minimum at $\Phi = 0$, i.e. $V'(0)=0$. In this case, the solution $\Phi = const = 0$ looks natural.  If $\Phi>0$, the system radiate waves of different kinds thus increasing the entropy of a thermostat. This process lasts until the energy is settled in a minimum, which is zero in our case.
	Such a picture is true if the gravity is absent. The latter leads to the gravitational instability like Jeans instability that is the reason for the large scale structure formation in our Universe. Scalar field instability in the framework of the Einstein gravity was discussed in \cite{Khlopov:1985jw} where the instability in the wavenumber range
	\begin{equation}
	0< k^2< k_J^2=4\sqrt{\pi G_N}m^2a_0,
	\end{equation}\label{Jeans}
	were found. Here $a_0$ is an initial amplitude of the field and $m$ is its mass. The final state could be compact objects made from the scalar field \cite{Carneiro:2018url}.

	Suppose that such compact object has been created within the extra dimensions provided that its density distribution along the $x$ coordinate is the uniform (i.e. the scalar field depends only on the extra coordinates $y$). Its stability may be supported by general arguments. Indeed, if such configuration decays, a 4-dimensional observer must detect a final state consisting of point-like defect of the scalar field distribution and massive scalar particles that have been instantly nucleated from the homogeneous state. Such a process is forbidden due to energy conservation. This argument for stability is not absolutely strict but reliable and we will keep it in mind postponing thorough study for the future.
	
	Numerical solutions of differential equations \eqref{eqn} depending on additional conditions and the scalar field acting in the extra space were studied in \cite{Gani:2014lka,Rubin:2015pqa} and we shortly reproduce them here. It was assumed that the metric of our 4-dim space is the Minkowski metric, $g_{4} = diag (1,-1,-1,-1)$.  The compact 2-dim manifold is supposed to be parameterized by the two spherical angles $y_1=\theta$ and $y_2=\phi$ $(0 \leq\theta \leq \pi, 0 \leq \phi < 2\pi)$.
	The choice of the extra space metric in \eqref{metric} is as follows
	\begin{equation}\label{metric2}
	g_{2,\theta\theta} = -r(\theta)^2;\quad g_{2,\phi \phi}= -r(\theta)^2 \sin^2(\theta).
	\end{equation}
	
	The system of equations \eqref{eqn} for two unknown functions acquires the form
	\begin{eqnarray}\label{eqn1}
&&	\partial^2_{\theta}R+\cot(\theta)\partial_{\theta}R=\frac12 r(\theta)^2\left[-R^2+\frac{c}{a}-\frac{m^2}{a}Y(\theta)^2\right], \\
&& 	\partial^2_{\theta}\phi(\theta)+\cot(\theta)\partial_{\theta}\phi(\theta)=m^2 r(\theta)^2 \phi(\theta). \label{eqn2}
	\end{eqnarray}
	All other equations in system \eqref{eqn} are reduced to trivial identities that is thoroughly analysed in \cite{2017JCAP...10..001B}.
	The Ricci scalar is expressed in terms of the radius $r(\theta)$
	\begin{equation}\label{Ricci}
	R=\frac{2}{r(\theta)^4\sin(\theta)}(-r'r\cos(\theta)+r^2 \sin(\theta)+r'^2 \sin(\theta)-\sin(\theta)rr''),
	\end{equation}
	where prime means $d/d\theta$.
	
	Let us fix the metric and the scalar field at the point $\theta =\pi$ 
	\begin{eqnarray}\label{ad1}
	&&R(\pi)=R_{\pi}, \quad R'(\pi)=0, \quad r(\pi)=\sqrt{2/R_{\pi}}, \quad r'(\pi)=0. \\
	&&\phi(\pi)=\phi_{\pi}, \phi'(\pi)=0. \label{ad2}
	\end{eqnarray}
			In the absence of matter, the extra metric is supposed to be the maximally symmetric, i.e. $R(\theta,\varphi)=const,\quad r=\sqrt{2/R}$.
	
\begin{figure}	
		\includegraphics[width=10cm]{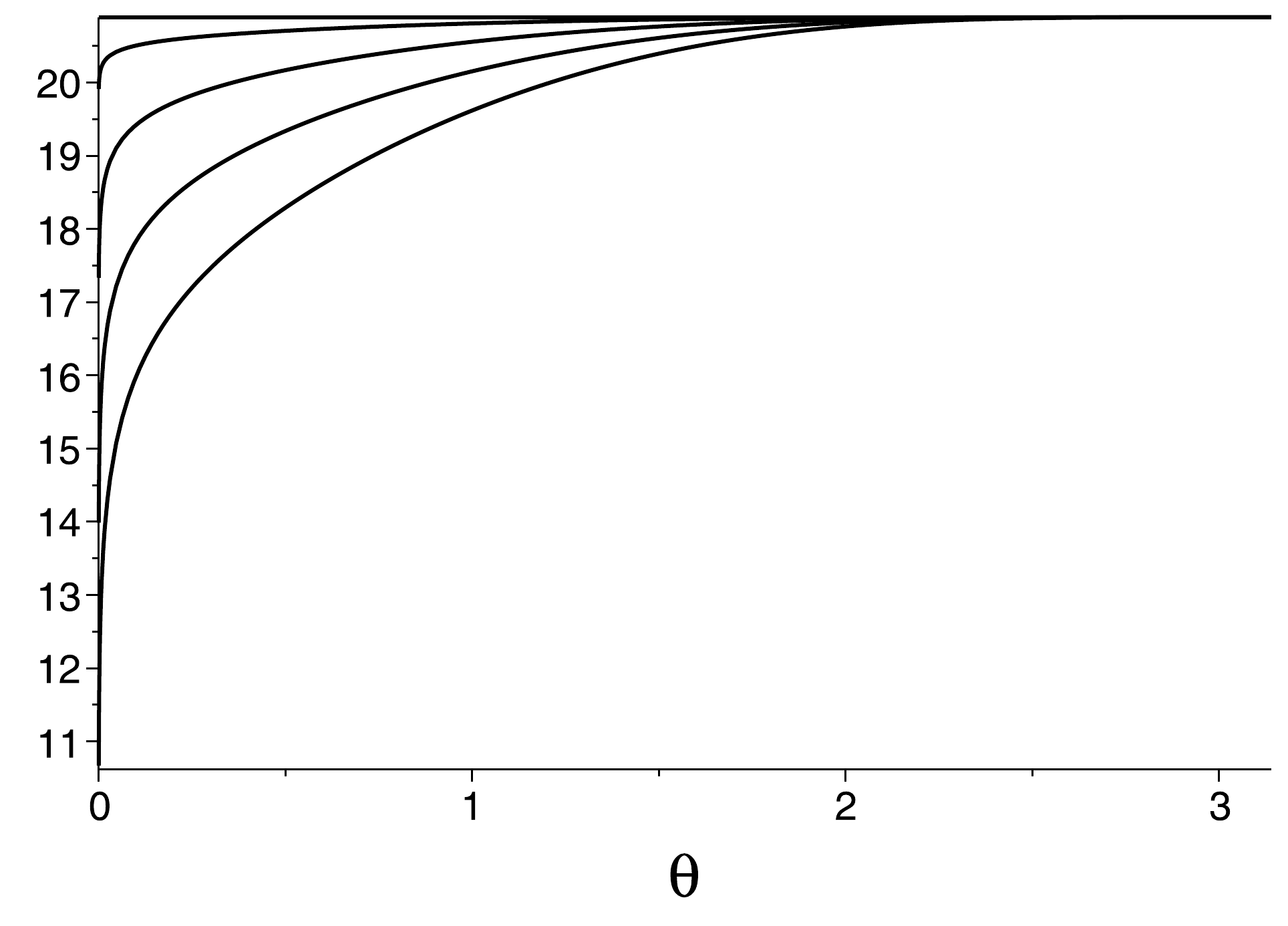} 
		\caption{Dependence of the 2-dim extra space radius $r(\theta)$ on the azimuthal angle $\theta$. The parameter values are $a = -100; b=1; c = -2.1\cdot 10^{-3}; m=0.01; m_D= 1$. 
		Additional conditions: $R_{\pi}=0.00458$, $\phi_{\pi}$ varies continuously within the interval $(0\div 2.28)$. Several points of this interval are taken: $\phi_{\pi}=0.57n, n=0,1,2,3,4$.	The more matter is placed in the extra space the more metric deviates from the sphere $r(\theta)=const$.
		}.	\label{MetricVsMatter}
	\end{figure}

	The system of equations \eqref{eqn1}, \eqref{eqn2} together with additional conditions \eqref{ad1}, \eqref{ad2} completely determine the form of extra space metric. The horizontal line in Fig.\ref{MetricVsMatter} (the scalar field is absent, $r(\theta)=const$) is the trivial solution to the system which coincides with our intuition and hence validates the applied method.   
	Nontrivial results for the extra metric were obtained for the nonzero value of the scalar field density within the extra space, see Fig.~\ref{MetricVsMatter} where numerical solutions to equations \eqref{eqn1}, \eqref{eqn2} are represented. The more scalar field is inserted into the extra dimensions, the deeper the well is formed. The scalar field density relates to the additional conditions $\phi_{\pi}$ at point $\theta =\pi$ which represent a set of the cardinality of the continuum. We conclude that the extra space metric continuously depends on the scalar field distribution in the extra space.
	
	As was discussed in the beginning of this Section, the Multiverse is the set of universes with accidental scalar field values. The latter is the reason of the accidental metrics of extra space in different parts of the Multiverse. Several examples of them are represented in Fig.\ref{MetricVsMatter}.
	
	\subsection{Matter induced branes and variation of 4-dim physical parameters}\label{param}
	
	In this subsection, the way to obtain effective 4-dim action for matter fields is discussed. To facilitate analysis, a trial scalar field $\chi$ is used as an example. Its action is written in the standard form
	\begin{eqnarray}\label{act0}
	&&S_{\chi}=\int d^D z \sqrt{|g_D|}\left[\frac12 \partial_{A} \chi g_{D}^{AB}\partial_{B} \chi - \frac{m_{\chi}^2}{2}\chi^2 \right].
	\end{eqnarray}
	Let us decompose the field around its classical part
	\begin{equation}\label{xY}
	\chi(x,y) =\chi_{cl}(y)+\delta\chi;\quad \delta\chi\equiv\sum_{k=1}^{\infty}\chi_k (x) Y_k(y)\ll \chi_{cl}(y),
	\end{equation}
	where $Y_k(y)$ are the orthonormal  eigenfunctions of the d'Alembert operator acting in the inhomogeneous extra space
	\begin{equation}\label{eqY1}
	\square_n Y_k(\theta)=l_k Y_k(y).
	\end{equation}
	The term $\chi_{cl}(y)$ is the solution to classical equation
	\begin{equation}
	\square_D \chi_{cl}(y)=\square_n \chi_{cl}(y)=-U'(\chi_{cl}(y))
	\end{equation}
	
	If we take into account the form of metric in the numerical example discussed above, the trial scalar field distribution can be written in the form
	\begin{equation}\label{cut}
	\chi_{cl}(\theta)= C\exp\{-m_{\chi}\int^{\theta} _0 d\theta' r (\theta')\}
	\end{equation}
	valid for not very small coordinate $\theta$, see discussion in \cite{Rubin:2015pqa}. The normalization multiplier $C$ defines the density of the scalar field distributed over the extra dimensions relates to an amount of the field stored in the extra dimensions from the beginning.

	Below, we limit ourselves by only first term in the sum \eqref{xY} so that
	\begin{equation}
	\delta\chi=\chi_0(x)Y_0(y), \quad \square_nY_0=0.
	\end{equation}
	After substitution \eqref{xY} into expression \eqref{act0} we get the following form of the effective 4-dim action for the gravity with the scalar field
	\begin{eqnarray}
	S_{\chi}=\frac{1}{2}\int d^4  x \sqrt{|g_4|}\left[ \frac{1}{2}\partial_{\mu}\chi_0(x)g^{\mu\nu}\partial_{\nu}\chi_0(x) - \frac{m^2}{2}\chi_0 (y)^2 - ... -\Lambda_{\chi} \right] , \label{Ssc}
	\end{eqnarray}
	where
	\begin{eqnarray}
	&&m^2=\int d^n y \sqrt{|g_n(y)|} \left[m^2_{\chi}Y_0(y)^2-\partial_aY_0 (y)g_n^{ab}(y)\partial_b  Y_0(y)\right] , \label{m2}\\
	&&\Lambda_{\chi}= \int d^n y \sqrt{|g_n(y)|} \left[\frac12 m^2_{\chi}\chi_{cl}^2 -\frac12 \partial_a\chi_{cl} (y)g_n^{ab}(y)\partial_b  \chi_{cl}(y)\right] \label{La}
	\end{eqnarray}
	The effective mass and $\Lambda_{\chi}$ term are the functions of the classical field distribution $\chi_{cl}(y)$ in the extra dimensions. The latter depends on an accidental conditions just after the D-dim manifold was formed.  Therefore, the mass of the scalar field $\chi_0$ (and the Lambda term) varies depending on initial conditions that have been realized at the early Universe. 
	
	The Higgs field is responsible for nonzero masses of the fermions and gauge bosons of the Standard Model. Hence, it is worth discussing the parameters of the Higgs potential and their possible variation. The simplest way is to introduce an interaction of the Higgs field and the field $\chi$ in the spirit of the moduli field approach, see \cite{Trigiante:2016mnt} and references therein. To this end, consider $D$-dim action as an example:
	\begin{eqnarray}
	&&S_H=\int d^4xd^ny\sqrt{g_4g_n}[\partial H^+\partial H +\mu^2(\chi)H^+H- \lambda(\chi)(H^+H)^2].
	\end{eqnarray}
	Here $\lambda(\chi)$ and $\mu^2(\chi)$ are arbitrary functions of the field $\chi(y)\simeq \chi_{cl}(y)$. Integration out the extra coordinates $y$ leads to the standard form of the action for the Higgs field with the parameters
	
	\begin{equation}
	\mu_{eff}^2=\int d^n y\mu^2(\chi_{cl}),\quad \lambda_{eff}=\int d^n y\lambda(\chi_{cl})
	\end{equation}
	where only zero mode of the Higgs field  $H=H(x)$ is taken into account.
	We have got the effective action
	\begin{eqnarray}
	&&S_{H,eff}=\int d^4xd^ny\sqrt{g_4g_n}[\partial H^+\partial H +\mu_{eff}^2H^+H- \lambda_{eff}(H^+H)^2].
	\end{eqnarray}
	with 4-dim parameters depending on the matter distribution in the extra dimensions. 
	
	The same can be said about the Planck mass and the Cosmological constant. Indeed, formulae  \eqref{MPl} and \eqref{Lambda} are easily converted to the expressions
	\begin{equation}\label{MPl2}
	M^2 _{P}=m_{D} ^{D-2}\int d^n y  \sqrt{|g_n(y)|}f' (R_n (y) )
	\end{equation}
	for the Planck mass and
	\begin{equation}\label{Lambda2}
	\Lambda \equiv -\frac{m_D ^{D-2}}{2M^2 _{P}}\int d^n y \sqrt{|g_n(y)|} [f(R_n(y)) + L_m (y)]
	\end{equation}
	for the cosmological $\Lambda$ term. Both the Planck mass and the $\Lambda$ term depend on a stationary geometry $g_{n,ab} (y)$ which are now functions of not only the Lagrangian parameters, but also on the accidental value of the initial scalar field density.

	The preliminary conclusion is as follows. The matter uniformly distributed in our 3-dim space can be a reason for the branes formation due to a nontrivial distribution of the matter within the extra dimensions. The brane properties depend not only on the Lagrangian parameters but also on the density of the matter (the scalar field $\phi$ in our case). The latter is a random value that is formed in the early Universe when the quantum fluctuations were important. One can conclude that a variety of branes with different properties can be formed in different spatial regions which could be a basis for the idea of the Multiverse. Therefore, this property could lead to the solution of the Fine-tuning enigma. Below we discuss this topic, bearing in mind the problem of the cosmological constant.
	
	\section{ Fine-tuning of the Lambda term and matter induced branes}
	
	The situation with the Lambda term remains intriguing \cite{Sahni:1999gb} despite two decades of discussion. Cosmological observations indicate that the current acceleration is described by the general relativity with the extremely small Cosmological Constant (CC). At the same time, the quantum fluctuations lead to the vacuum energy density which is in many orders of magnitude higher than the observed value of the CC. There are many models elaborated to explain the smallness of the $\Lambda$ term, see e.g. \cite{Yurov:2005zn, Garriga:2000cv}. General discussion on the subject can be found in \cite{Weinberg:1987dv,Sahni:1999gb,Loeb:2006en,Wetterich:2008sx,Bousso:2000xa,Brown:2013fba,Burgess:2013ara}.
	
	The role of quantum corrections is not clarified up to now. The quantum corrections are of the order of cutting parameter which is compatible with a highest energy scale of a specific model. Hence natural values of physical parameters defined at this energy scale are of the same order of magnitude as this highest energy scale and it is not clear how to neutralize them except by a strong parameter selection. We have to admit that the observed Lambda term value is hardly be explained in terms of the physical parameters determined at low energies. The problem is deepened because if this value were several times larger, intelligent life would be absent. This represents the particular case of the fine-tuning problem.
	
	The question "How the physical parameters acquire the observable values?"\ divides the physical community into two groups. The first one does not bother with questions of such kind. They are interested in the study of physical laws that explain experimental facts.  This point of view is quite firm but slightly inconsistent. Indeed, there is the experimental fact of fine-tuning of the physical parameters necessary for the existence of intelligent life.
	The range of the parameter values is very narrow and like any observed phenomena, it must be explained. This is the reason for the second group of physicists to make efforts in answering this question.
	
	The first step has been done decades ago when the Anthropic principle was proclaimed: "there are a lot of different patches (or universes) with different properties and the life originates in universes with appropriate conditions". The immediate question is formulated as follows: What is the mechanism for the creation of a variety of universes (Multiverse) with different properties? As we will see, an attempt to answer this question consists of several ideas that deserve further development. The Anthropic principle is not the solution to the fine-tuning problem but the small step forward.
	
	The string theory is the well-known idea supplying us with the multiverse - the landscape in its terms \cite{Susskind:2003kw}. Unfortunately, this approach has a weakness. Indeed, even if a number of final states is as huge as $10^{500}$ in the string theory, they could be distributed non uniformly in parameter space and there is no certainty that the necessary physical parameters can be realized.
	This shortcoming can be eliminated if the set of low energy parameters has the cardinality of the continuum. This relates to the discussion made above. The branes induced by matter depend on accidental values of the initial energy density of matter produced by the quantum fluctuations. Therefore a set of such branes has the cardinality of the continuum. Evident logical chain is: \textit{continuum set of initial metrics $\rightarrow$ continuum set of final metrics $\rightarrow$ continuum set of the $\Lambda$-terms.}

	The picture looks as follows. In the spirit of the inflationary scenario, quantum fluctuations at high energies produce a huge variety of space volumes characterized by different energy density and hence by branes with different properties. This is the reason for the formation of different cosmological constants within such volumes. 
	
\begin{figure}
		\includegraphics[width=10cm]{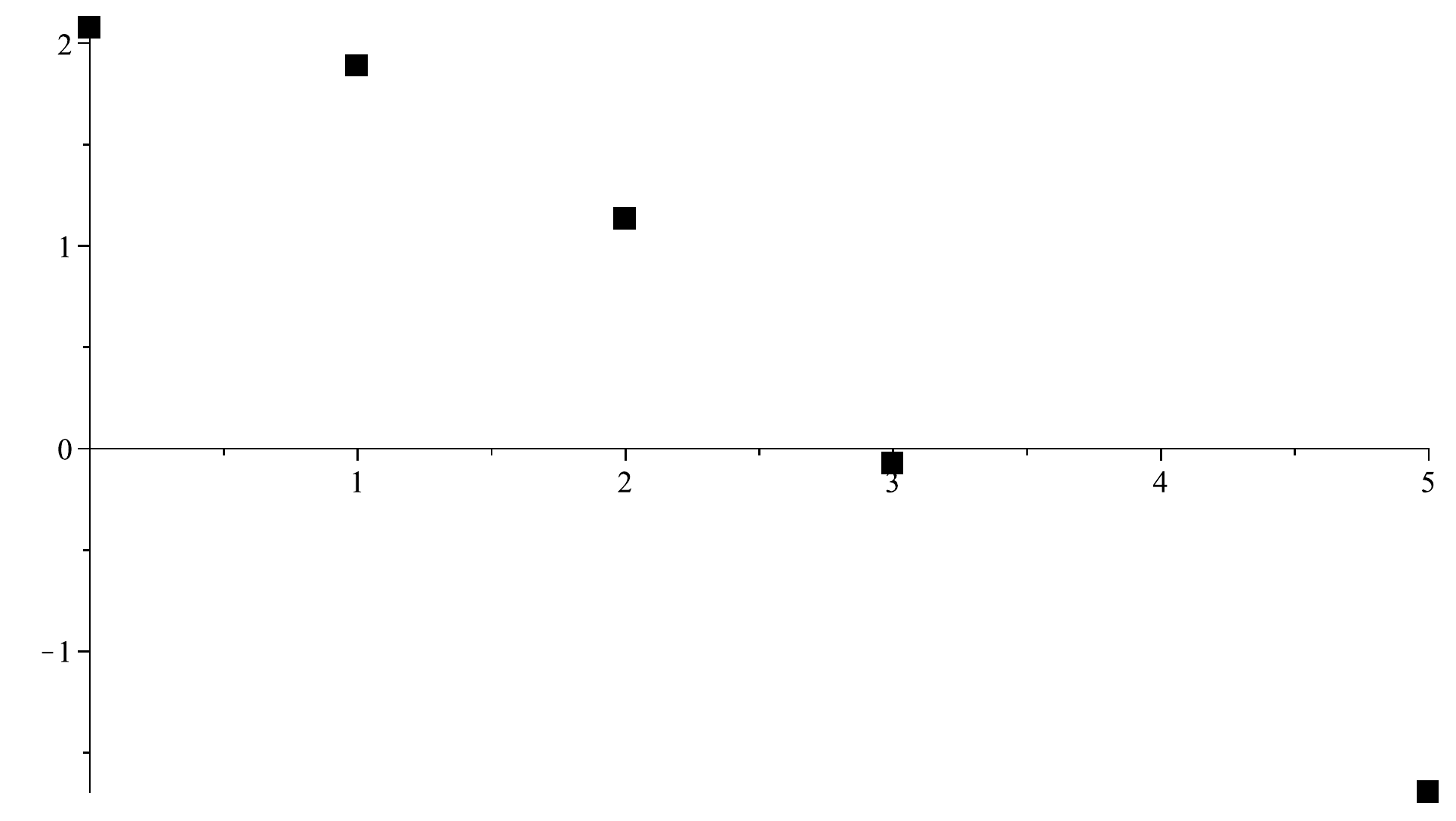} 
		\caption{Dependence of the $\Lambda$-term (arbitrary units), see  \eqref{Lambda2} on the discrete set of the additional conditions $\phi_i(\theta=\pi),\quad i=1,...,5$. Each point relates to the specific curve in Fig.\ref{MetricVsMatter}. The cosmological constant varies from negative to positive values. At the same time, their set has the cardinality of the continuum. Therefore, we sure can find such additional conditions for which the $\Lambda$-term is arbitrary close to zero.
		}
		\label{Lambda}
	\end{figure}

	According to formula \eqref{Lambda2}, the value of the Lambda depends on the scalar field distribution along the extra dimensional coordinates. One can see from Fig.\ref{Lambda} that the cosmological constant varies from negative to positive values due to variation of the matter distribution. The latter relates to the additional condition that fixes the scalar field at the boundary $\phi(\theta=\pi)$. The universes differ from each other due to the initial distribution of the matter along the extra dimensions. The problem of the $\Lambda$-term smallness is reduced now to the question "does this set contain the term $\Lambda =0$?". Fig. \ref{Lambda} gives the positive answer to the question if one keeps in mind that the additional conditions varies continuously. 
	In particular, there exists a set of universes with such initial matter distributions that gives the cosmological $\Lambda$ terms be arbitrarily close to zero.
	
	\section{Conclusion}
	
	There are at least two general problems of the cosmology that worth discussing - the Hierarchy problem and the Fine-tuning one. 
	It seems that a multidimensional paradigm allows us to solve the first puzzle, while the exact adjustment of the physical parameters remains unresolved. The perspective way to solve it is an elaboration of mechanism of the Multiverse formation containing a continuum set of different Universes. The mechanism of such sort is discussed in this article.
	
	The main point is the use of matter to obtain a non-trivial metric of extra space. The 4-dim analogy can be useful. Indeed, the formation of compact dense objects due to the Jeans instability leads to the formation of a variety of objects, the mass of which depends on an initial matter distribution. The same process could take place in the extra dimensions where compact objects - branes - are formed under the influence of matter.
	
	Each universe belonging to the Multiverse is described by the specific distribution of matter and hence by specific extra space metric. This leads to the formation of causally disconnected regions endowed by branes that differ with each other. Therefore the physical parameters in such volumes are also different as was discussed above. 
	
	Initial conditions form a continuous set. Hence, the extra space metrics also form a set of the cardinality of the continuum. The low energy physical parameters depend on the extra space metrics and hence represent a continuous set as well. This means that those space areas where the Lambda term has the observable value do exist thus providing the basis for the Anthropic argument. 
	
	The quantum fluctuations seem to destroy the analysis made on the classical level. This problem is discussed in \cite{Rubin:2016ude} where it was shown that the situation looks solvable in the framework of the effective field theory. Nevertheless, thorough renormgoup analysis has to be performed in the future.
	
	The discussion in this article shows that the matter distribution within the extra space is a promising way to describe the fine-tuning of the physical parameters. The thick branes become the substantial tool for study wide class of topical questions. The idea could be applied to explain e.g. the number of particle generations, the inflation, the primordial black holes formation and so on, see review \cite{Liu:2017gcn}.
	
	\section{Acknowledgement}
	The work was supported by the Ministry of Education and Science of the Russian Federation, MEPhI Academic Excellence Project (contract N~02.a03.21.0005, 27.08.2013).
	The work was also supported by the grant RFBR N~19-02-00930 and is performed according to the Russian Government Program of Competitive Growth of Kazan Federal University. 
	

\end{document}